\documentclass[prl,twocolumn,showpacs,aps,superscriptaddress]{revtex4-1}

\usepackage{graphicx}
\usepackage{dcolumn}
\usepackage{bm}
\usepackage{amsthm}
\usepackage{dsfont}
\usepackage{array}
\usepackage{amsmath}
\usepackage{mathbbol}
\usepackage{epic}
\usepackage{calc}
\usepackage{amsfonts}
\usepackage{amssymb}

\newtheorem*{theorem*}{Theorem}

\newtheorem*{lemma*}{Lemma}

\newtheorem*{propo*}{Proposition}

\DeclareMathOperator{\Tr}{Tr}
\providecommand{\openone}{\leavevmode\hbox{\small1\kern-3.8pt\normalsize1}}

\begin{document}
\title{Quantum steering of Gaussian states via non-Gaussian measurements}
\author{Se-Wan Ji}
\affiliation{Department of Physics, Texas A$\&$M University at Qatar University, PO Box 23784, Doha, Qatar}
\author{Jaehak Lee}
\affiliation{Department of Physics, Texas A$\&$M University at Qatar University, PO Box 23784, Doha, Qatar}
\author{Jiyong Park}
\affiliation{Department of Physics, Texas A$\&$M University at Qatar University, PO Box 23784, Doha, Qatar}
\author{Hyunchul Nha}
\affiliation{Department of Physics, Texas A$\&$M University at Qatar University, PO Box 23784, Doha, Qatar}
\begin{abstract}
Quantum steering---a strong correlation to be verified even when one party or its measuring device is fully untrusted---not only provides a profound insight into quantum physics but also offers a crucial basis for practical applications. For continuous-variable (CV) systems, Gaussian states among others have been extensively studied, however, mostly confined to Gaussian measurements. While the fulfillment of Gaussian criterion is sufficient to detect CV steering, whether it is also necessary for Gaussian states is a question of fundamental importance in many contexts. This critically questions the validity of characterizations established only under Gaussian measurements like the quantification of steering and the monogamy relations.
Here, we introduce a formalism based on local uncertainty relations of non-Gaussian measurements, which is shown to manifest quantum steering of some Gaussian states that Gaussian criterion fails to detect. To this aim, we look into Gaussian states of practical relevance, i.e. two-mode squeezed states under a lossy and an amplifying Gaussian channel. Our finding significantly modifies the characteristics of Gaussian-state steering so far established such as monogamy relations and one-way steering under Gaussian measurements, thus opening a new direction for critical studies beyond Gaussian regime.   
\end{abstract}
\maketitle

\section*{Introduction}

Quantum correlations, which do not admit classical descriptions, profoundly distinguish quantum physics from classical physics and also provide key resources for quantum information processing \cite{Nielsen}. Under the classification of quantum correlations, the quantum non-locality is the strongest form of correlation \cite{Brunner14} beyond any classical local hidden variable (LHV) models \cite{Bell64}, which can be used e.g. to achieve the unconditional security of key distribution (QKD) in a device-independent manner \cite{Acin07, Pironio09, Masanes11}. In LHV model, the joint probability for outcomes $a$ and $b$ of local measurements $A$ and $B$, respectively, is of the form 
$P_{LHV}\left( a,b | A,B \right) =\sum_{\lambda} p \left( \lambda \right) P\left( a|A, \lambda \right)P\left(b|B,\lambda \right),$
with a hidden variable $\lambda$ distributed according to the probability $ p\left( \lambda \right)$. Here, for a given $\lambda$, the local probability distributions are realized independently of each other. Notably, any local statistics are allowed even beyond quantum ones in the LHV models. On the other hand, if the conditional probabilities 
are restricted only to quantum statistics, i.e.  
$P_Q \left( a,b | A, B\right) = \sum_{\lambda} p\left( \lambda \right) P_Q\left( a| A, \rho_A^{\lambda} \right) P_Q \left( b|B, \rho_B^{\lambda} \right),$
where the subscript $Q$ refers to quantum statistics, the model describes quantum separability, the violation of which indicates quantum entanglement---a resource, e.g. for exponential speed-up over classical computation \cite{Nielsen,Jozsa03}.
Recently, an intermediate form of correlation has been rigorously defined as quantum steering \cite{Wiseman07}, which has become a topic of growing interest over the past decade \cite{Wiseman07, Cavalcanti09, Branciard12, Handchen12, one-way, Midgley10, Olsen13, HeR13, HeF14, Tan15, Smith12, Wittman12, Bennett12, Quintino14, Uola14, Bowles14, Piani15}. In the so-called local hidden state (LHS) model, the correlation is represented by the form
\begin{equation}
\label{nonsteer1}
P_{LHS}\left( a,b |A,B \right)= \sum_{\lambda}p \left( \lambda \right) P_Q\left( a| A, \lambda \right) P\left( b| B, \rho_{\lambda} \right),
\end{equation}
i.e. only the party $A$ is restricted to quantum statistics whereas the party $B$ is not. If the above approach fails to explain a certain correlation of a given state, the state is called steerable from $B$ to $A$. Not to mention its fundamental interest, it can have some important applications. A steering test is to certify quantum correlation even if Bob or his device is fully untrusted. In this sense, quantum steering may offer a crucial practical basis, e.g. for one-sided device independent cryptography \cite{Branciard12}, and was also shown to be useful for quantum sub-channel discrimination \cite{Piani15}. \

Quantum steering has been intensively studied for continuous variables (CVs) \cite{Braunstein05} as well as discrete variables. In fact, the first steering criterion was developed by M. Reid for CVs \cite{Reid89, Reid09} and a CV system is an attractive platform making an excellent candidate for quantum information processor. A wide range of quantum systems can process quantum information on CVs including light fields, collective spins of atomic ensembles, and motional degrees of trapped ions, Bose-Einstein condensates, mechanical oscillators \cite{Cerf}. An important regime to address CV systems both theoretically and experimentally is the one dealing with Gaussian states, Gaussian measurements and Gaussian operations \cite{Weedbrook12}. 
Although the Gaussian regime has many practical advantages due to its experimental feasibility, it has also been known that non-Gaussian operations and measurements are essential for some tasks such as CV nonlocality test \cite{Nha04, Patron04, Bell64} and universal quantum computation \cite{Lloyd99, Bartlett02}. On the other hand, there are also cases that Gaussian states and operations provide an optimal solution, e.g. classical capacity under Gaussian channels \cite{Mari14}, quantification of entanglement \cite{Giedke03} and quantum discord \cite{Adesso10, Pirandola14,Ollivier01}. These examples indicate that Gaussian regime can be sufficient for the characterization of certain quantum correlations of Gaussian states \cite{Giedke03, Adesso10, Pirandola14}. In particular, Gaussian criterion is both sufficient and necessary to detect quantum entanglement for Gaussian states \cite{Duan00,Simon00}.\\

So far, most of the studies on the steering of Gaussian states have been confined to Gaussian measurements, which established numerous characteristics of Gaussian-state steering \cite{Wiseman07, Cavalcanti09, Reid89, Reid13, He13, Adesso15, Nha15}. One remarkable example is a recent experiment that showed one-way steering of a two-mode Gaussian state, i.e. Alice steers Bob but Bob cannot steer Alice, under homodyne measurements \cite{Handchen12}. Another notable property is a strict monogamy relation identified under Gaussian measurements, i.e. Eve cannot steer Bob if Alice steers Bob \cite{Reid13, Nha15}, which can be a crucial basis for secure communication, e.g. in one-sided device independent cryptography \cite{Walk14}. 
On the other hand, it is very important to answer a critical question whether these properties under the restriction of Gaussian measurements can generally be true beyond the Gaussian regime. For instance, a special class of non-Gaussian measurements, i.e., higher-order quadrature amplitudes, was employed to study steerability of Gaussian states \cite{Nha15}, which did not show any better performance than Gaussian measurements. Kogias and Adesso further conjectured that for all two-mode Gaussian states, Gaussian measurements are optimal \cite{Kogias15}. It is a conjecture of both fundamental and practical importance to prove or disprove.


In this Article we demonstrate that there exist two-mode Gaussian states for which Gaussian measurements cannot manifest steering, but non-Gaussian measurements can. For this purpose, we employ a formulation of steering criterion based on local uncertainty relations involving non-Gaussian measurements \cite{Ji15}. We study steerability of mixed Gaussian states, specifically, Gaussian states under a lossy and an amplifying channel that represent typical noisy environments. Our counter-examples to the aforementioned conjecture imply the breakdown of monogamy relations and one-way steerability emerging under the restriction to Gaussian measurements and point out a critical need for more studies beyond Gaussian regime. \\

\section*{Results}
 
\subsection*{Gaussian vs. non-Gaussian steering criterion}
Let us define two orthogonal quadrature amplitudes $ {\hat X}_i=\frac{1}{\sqrt{2}}\left( {\hat a}_i + {\hat a}_i^{\dagger} \right)$, ${\hat P}_i=\frac{1}{\sqrt{2}i} \left( {\hat a}_i-{\hat a}_i^{\dagger} \right)$, where ${\hat a}_i$ and ${\hat a}_i^{\dagger}$ are the annihilation and the creation operator for $i$-th mode. The covariance matrix $ \gamma_{AB}$ of a bipartite $\left(M + N\right)$-mode Gaussian state $\rho_{AB}$ is then given by $\gamma_{AB}=\left( \begin{array}{cc}
\gamma_{A} & C \\
C^{T} & \gamma_{B} \\
\end{array} \right)$ whose elements are
\begin{equation}
\label{gcova}
\gamma_{AB}^{ij} = \langle \Delta {\hat r}_i \Delta {\hat r}_j +\Delta {\hat r}_j \Delta {\hat r}_i\rangle-2\langle\Delta {\hat r}_i \rangle \langle \Delta {\hat r}_j \rangle ,
\end{equation}
with ${\hat r}_i \in \left\{{\hat X}_1,{\hat P}_1,{\hat X}_2,{\hat P}_2,\cdots,{\hat X}_{M+N},{\hat P}_{M+N} \right\}$ and $\Delta {\hat r}_k={\hat r}_k-\langle {\hat r}_k \rangle$. $ \gamma_{A} $ and $\gamma_{B}$ are $2M \times 2M $, $2N \times 2N $ real symmetric positive matrices representing local statistics of $M$- and $N$-modes, respectively. $C$ is a $ 2M \times 2N $ real matrix representing correlation between two subsystems. Due to the uncertainty principle, the covariance matrix must satisfy
$\gamma_{AB} \pm i \Omega_{AB} \geq 0$,
where $ \Omega_{AB} =\oplus_{i=1}^{M+N} \left( \begin{array}{cc}
0 & 1 \\
-1 & 0 \\
\end{array} \right) $ is the sympletic form \cite{Simon}. For the case of a two-mode Gaussian state, the covariance matrix can always be brought to a standard form by local Gaussian operations \cite{Duan00, Simon00} 
\begin{equation}
\label{scova}
\gamma _{AB}  = \left( \begin{array}{cc}
 \gamma_A & C \\ 
 C^T & \gamma_B \\ 
 \end{array} \right) = \left( \begin{array}{cccc}
 a & 0 & c_1 & 0 \\ 
 0 & a & 0 & - c_2  \\ 
 c_1 & 0 & b & 0 \\ 
 0 & - c_2 & 0 & b \\ 
 \end{array} \right) .
\end{equation} 
Without loss of generality, we may set $ c_1 \geq \left| c _2 \right| \geq 0 $. \\
Under Gaussian measurements, a Gaussian state $ \rho_{AB} $ is non-steerable from $B$ to $A$ iff
\begin{equation}
\label{GC}
{\gamma _{AB}} + i{\Omega _A}\oplus{{\bf 0}_B} \ge 0,
\end{equation}
where $ {\bf 0}_{B} $ is a $ 2N \times 2N $ null matrix \cite{Wiseman07}.

We now introduce our non-Gaussian steering criterion based on a set of local orthogonal observables, which we use to detect the steerability of two-mode Gaussian states. 

{\bf Lemma}: 
Let us choose a collection of $n^2$ observables $\left\{ A^{(n)} \right\} = \left\{ \lambda_k,\, \lambda_{kl}^{\pm} \right\}$ $\left( k,l=0,1,...,n-1 \right) $ where
\begin{eqnarray}
\lambda_k&=& |k\rangle \langle k |\\
\lambda_{kl}^{+}&=&\frac{|k \rangle \langle l| +|l \rangle \langle k| }{\sqrt{2}}\,\, \left( k < l \right), \\
\lambda_{kl}^{-}&=&\frac{|k \rangle \langle l| -|l \rangle \langle k| }{\sqrt{2}i}\,\, \left( k <l \right). 
\end{eqnarray}
Here $ |k \rangle $ refers to a Fock state of number $k$. 
Note that each of $\lambda_k$ and $\lambda_{kl}^{\pm}$ represents a complete projective measurement individually. 
That is, $\lambda_k$ represents a two-outcome projective measurement, assigning $+1$ if the tested state is detected in $|k\rangle$ and $0$ otherwise.  On the other hand, $\lambda_{kl}^{\pm}$ represents a three-outcome projective measurement, assigning $\pm1$ if detected in the two eigenstates of $\lambda_{kl}$ and 0 otherwise.  These observables satisfy the orthogonal relations $ \Tr\left( A_i^{(n)} A_j^{(n)} \right) =\delta_{ij}$ and the sum of variances must satisfy the uncertainty relation,
\begin{equation}
\label{LUR1}
\sum_j \delta^2\left( A_j^{(n)} \right) \geq \left(n-1 \right) \langle \openone_n \rangle,
\end{equation}
where $\delta^2\left(A_j^{(n)}\right)= \langle \left(A_j^{(n)}\right)^2 \rangle-\langle A_j^{(n)} \rangle^2$ represents each variance and $ \openone_n = \sum_{k=0}^{n-1} |k \rangle \langle k | $ the projection to the Hilbert space of truncated $n$-levels. The proof of Lemma is given in Supplementary Information.   \\ 
\\

Now consider an orthogonal $n^2 \times n^2$ matrix $ O_n $  in the truncated space, satisfying $ O_n^T O_n= O_n O^T_n = \openone_n$. Then the set of observables $ \left\{ \tilde{A}_j^{(n)} \right\}$ under the transformation $\tilde{A}_j^{(n)}  = \sum_l O_{jl} A_l^{(n)}$ also satisfies the uncertainty relation in equation (\ref{LUR1}) since $ \sum_{j} \left( \tilde{A}_j^{(n)} \right)^2 = \sum_{j,l,l'} O^T_{lj}O_{jl'} A_l^{(n)}A_{l'}^{(n)}=\sum_l \left(A_l^{(n)}\right)^2 =n \openone_n$ and $ \sum_j \langle \tilde{A}_j^{(n)} \rangle^2 = \sum_{j,l,l'} O^T_{lj}O_{jl'}\langle A_l^{(n)} \rangle\langle A_{l'}^{(n)} \rangle = \sum_l \langle A_l^{(n)} \rangle^2 $. This invariance property will be used later. \\
\\
We now obtain a non-steering condition based on these orthogonal observables.\\
\\
{\bf Theorem}: 
If a two-mode quantum state $\rho_{AB}$ is non-steerable from B to A, it must satisfy the inequality
\begin{equation}
\label{LUR-steer}
\sum_j \delta^2 \left(  A_j^{(n)} \otimes \openone + g \openone \otimes  B_j^{(n')} \right) \geq \left(n-1 \right)\langle \openone_n^{A} \rangle,
\end{equation}
for any real $g$, where the observables $ \left\{ A_j^{(n)} \right\} $ satisfy the uncertainty relation in Eq. (\ref{LUR1}). Its proof is given in Supplementary Information. \\
\\ 
We next introduce a useful criterion originating from the inequality (\ref{LUR-steer}) that can be readily computable.\\
\\
{\bf Proposition}: 
If a two-mode state $ \rho_{AB}$ is non-steerable from $B$ to $A$, the correlation matrix $ C_{nn'}^{TLOOs}$ with elements $\left( C_{nn'}^{TLOOs} \right)_{ij}\equiv\langle A_i^{(n)} \otimes B_j^{(n')} \rangle-\langle A_i^{(n)} \rangle\langle B_j^{(n')} \rangle$ constructed with $n$- and $n'$-level truncated local orthogonal observables (TLOOs) must satisfy 
($ \|\,\cdot\|_{tr}$: trace norm)
\begin{eqnarray}
\label{steering-tr}
&&\| C_{nn'}^{TLOOs} \|_{tr} \nonumber\\
&&\leq \sqrt{ \left( \langle \openone_n^{A} \rangle -\sum_{j=1}^{n} \langle A_j^{(n)} \rangle^2 \right) \left( n' \langle \openone_{n'}^B \rangle - \sum_{j=1}^{n'} \langle B_j^{(n')} \rangle^2 \right)}.\nonumber\\
\end{eqnarray}
The proof of Proposition in Supplementary Information clearly shows that if a given state violates the inequality (\ref{steering-tr}), it is steerable from B to A, and also that there exist a set of TLOOs that violates the inequality (\ref{LUR-steer}) as well. We give some details on how to calculate the expectation values of TLOOs for a given state in Supplementary Information.                  

\begin{figure}[!t]
  \includegraphics[width=0.5\textwidth]{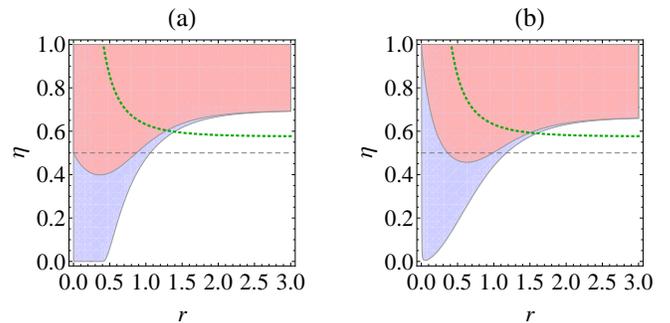}
    \centering
\caption{Detection of steerability for a lossy Gaussian state based on the criterion in equation (\ref{steering-tr}) using (a) $2$-level TLOOs and (b) 3-level TLOOs, respectively. Initially, a two-mode squeezed state with squeezing parameter $r$ is prepared and only the mode $B$ undergoes a lossy channel with transmittance rate $\eta$. The shaded red (blue) regions in the figures manifest a successful detection of steering from $B$ to $A$ (from $A$ to $B$). The dashed gray straight line $\eta=1/2$ is the bound under which Gaussian steering criterion cannot detect steering from $B$ to $A$. Regions above dotted green curves represent the two-mode Gaussian states with which a positive key extraction is possible using CV OSDIQKD  of Ref. \cite{Walk14}.     }
\label{lossfig}
\end{figure}
  
\subsection*{Application to Gaussian states}
Let us consider a two-mode squeezed-vacuum (TMSV) state as an initial state with its covariance matrix $\gamma_{AB}^{TMSV}$ given by $ a =b=\cosh{2r}$ and $c_1=c_2=\sinh{2r}$ in equation (\ref{scova}). ($r$: squeezing parameter). If $r>0$, this state is always steerable in both ways with Gaussian measurements using equation (\ref{GC}). 
Now suppose that the mode $B$ undergoes a vacuum noisy channel with transmission rate $\eta$. Then the output state has a covariance matrix \cite{Nha15},
\begin{eqnarray}
\label{losscova}
&&\gamma_{AB}^{TMSV}\rightarrow\gamma_{AB'}^{TMSV}=\nonumber\\
&&\left( \begin{array}{cccc}
\cosh{2r} & 0 & \sqrt{\eta}\sinh{2r} & 0 \\
0 & \cosh{2r} & 0 & -\sqrt{\eta}\sinh{2r} \\
\sqrt{\eta}\sinh{2r} & 0 & \eta \cosh{2r}+1-\eta & 0 \\
0 & -\sqrt{\eta}\sinh{2r} & 0 & \eta\cosh{2r}+1-\eta \\
\end{array} \right).\nonumber
\end{eqnarray}
If $\eta>\frac{1}{2}$, one can readily check that the Gaussian criterion detects steering from $B$ to $A$ for all states regardless of squeezing. However, if the transmission rate is below $\eta=\frac{1}{2}$, steering is impossible from $B$ to $A$ under Gaussian measurements \cite{Nha15}.  On the other hand, the steering from $A$ to $B$ is always possible regardless of $\eta>0$ for a nonzero squeezing.
In contrast, as we show in Fig. \ref{lossfig}, one can detect steerability of noisy Gaussian states from $B$ to $A$ even below $\eta = 1/2$ using non-Gaussian measurements based on 2-level [Fig. 1 (a)] and 3-level [Fig. 1 (b)] TLOOs. For this calculation, we need the density matrix elements in Fock basis, $ \rho_{m_1m_2n_1n_2}\equiv{\rm Tr} \{\rho | m_1 \rangle_A \langle n_1| \otimes | m_2 \rangle_B  \langle n_2 | \}$, which are given in Methods for completeness.

Our motivation for the choice of TLOOs in low-photon numbers is rather natural. For a low transmission rate $\eta$, the mode $B$ resides in the Fock space of low photon numbers and this is particularly true for a small initial squeezing $r$. It is then expected that the information on correlation exists largely in the low-photon Fock space. For the case of 2-level TLOOs in Fig. 1 (a), the detection range under our steering test, for which Gaussian criterion fails, turns out to be $0<r<0.869$ (red shaded region below dashed line). On the other hand, for the case of 3-level TLOOs in Fig. 1 (b), it turns out to be $0.364<r<0.987$ (red shaded region below dashed line). One might then be interested to see if a truncated TMSS with a proper normalization in a genuinely low-dimensional Hilbert space can also show steering in a similar way. We find that it does not detect steerability beyond Gaussian criterion. In a sense, our TLOOs constructed with low-dimensional Fock states obtain a coarse-grained information on higher-order terms, not completely ignoring them, in an on-off fasion. (See the paragraph below equation (7).) 

Other Gaussian states of experimental relevance are TMSVs under an amplifying channel. Let mode $B$ undergo the amplification channel with a gain factor $ G \geq 1$, i.e. ${\hat b}_{\rm out}=\sqrt{G} {\hat b}_{\rm out}+ \sqrt{G-1}{\hat v}$, where ${\hat v}$ represents a vacuum noise. Then the output covariance matrix is given by $ a=\cosh{2r}$, $c_1=c_2=\sqrt{G}\sinh{2r}$, and $ b=G\cosh{2r}+G-1$ in equation (\ref{scova}). In this case, steering is possible from $A$ to $B$ using Gaussian criterion in the range of $ 1 \leq G < \frac{2\cosh{2r}}{\cosh{2r}+1}$, whereas it is so from $B$ to $A$ in all ranges of $G$ for a nonzero $r$. On the other hand, if we choose $2$-level TLOOs in each party and test the violation of our steering criterion in Eq. (\ref{steering-tr}), we find that the states with $ G = \frac{2\cosh{2r}}{\cosh{2r}+1}+\epsilon $ are detected where $ 0 \leq \epsilon \lesssim 0.05$ for $ 0< r \lesssim 0.65 $, where a nonzero $\epsilon$ indicates that there are some amplified Gaussian states the steering of which is detected not via Gaussian criterion but via our non-Gaussian measurements.

\begin{figure}[!t]
  \centering
  \includegraphics[width=0.4\textwidth]{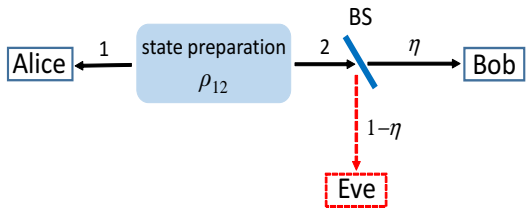}
\caption{Modeling a lossy channel by a beamsplitter (BS) of transmittance $\eta$.}
\label{model}
\end{figure}

\section*{Discussion}
Aside from its fundamental interest, our result can have some practical implications as well. First, it has been known that there exists a strict monogamous property of steering under Gaussian measurements \cite{Reid13, Nha15}. That is, if Bob can steer Alice's system, Eve cannot steer it simultaneously. In contrast, our result provides a clear counter-example to this monogamy relation. For example, a loss channel can be modeled by a beamsplitter as shown in Fig. 2. Then, if Bob possesses a field of fraction $\eta$, Eve takes the remaining fraction $1-\eta$. In the case of e.g. $\eta=0.55$, Bob can steer Alice's system via Gaussian measurements. At the same time, however, Eve can also steer Alice's system via non-Gaussian measurements because it corresponds to the case of $1-\eta=0.45$ at which steering is possible as shown in Fig. 1. Therefore, there occurs a possibility of simultaneous steering if not restricted to the same Gaussian measurements. A similar argument can be given to the case of amplified states.

In a related context, one may wonder if the above breakdown of monogamy relation can have implication on the security of one-sided device independent quantum key distribution (OSDIQKD). 
Basically, steering is a necessary condition to establish a positive key rate for OSDIQKD. If the monogamy relation does not hold, i.e. a simultaneous steering is possible, the security can be potentially compromised. However, importantly, it is not an arbitrary form of steering but a specific one that matters for a given protocol. For example, the CV OSDIQKD scheme of Ref. \cite{Walk14} extracts  keys based on specific observables at Alice’s station, which are two orthogonal quadratures $\hat X$ and $\hat P$ (Fig.1). If only two observables are considered for the purpose of steering in the trusted party, a simultaneous steering is impossible, which was first shown by Reid \cite{Reid13}. This monogamy argument is even valid regardless of the type of states, whether Gaussian or non-Gaussian. Therefore, if Alice can establish a positive key rate with Bob, Eve cannot. 

Second, the interesting phenomenon of asymmetric steering, i.e. Alice steers Bob but Bob cannot steer Alice, which was so far investigated under the restriction to Gaussian measurements \cite{Handchen12,one-way}, must be rigorously reassessed. For example, the recent experiment \cite{Handchen12} studied the case of Gaussian states under a lossy channel and our result demonstrates that the transmission rate $\eta=1/2$ is not a critical value for the one-way steering.

In summary we showed that there exist Gaussian states the steering of which Gaussian measurements cannot detect but non-Gaussian measurements can. To this aim, we introduced a criterion based on local orthogonal observables and their uncertainty relations in a truncated Hilbert space.  We have applied this criterion to the case of TMSV under a lossy and an amplifying channel and found that Gaussian measurements are not always optimal to demonstrate steering of Gaussian states. Our result implies that the steering properties known under the restriction to Gaussian measurements must be rigorously reassessed. For example, the important properties such as the strict monogamy relation and asymmetric steering break down beyond Gaussian regime. 

Our investigation clearly indicates that we must pursue more studies to completely understand the characteristics of quantum steering even for the restricted set of Gaussian states. We hope our finding here could provide some useful insights into future studies beyond Gaussian measurements and operations.

{\it Remarks}: Upon completion of this work \cite{arxiv}, we became aware of a related work in \cite{Pryde} that employs pseudo-spin observables to show non-optimality of Gaussian measurements for steering of Gaussian states. We here briefly compare their method with ours particularly in detecting Gaussian states under a loss channel. As shown in Fig. 3, the method of Ref. \cite{Pryde} detects steering at a lower level of transmittance $\eta$ for the squeezing range $0<r<0.746$ ($0<r<0.743$) than our 2 (3)-level TLOO criterion. However, it does not detect steering below $\eta=0.5$, which is the case of interest as Gaussian criterion fails, if the squeezing level is rather high as $r>0.81$. On the other hand, our method detects steering in the range of $0<r<0.869$ and $0.364<r<0.987$ using 2- and 3-level TLOO criterion, respectively, below $\eta=0.5$. Thus, the red and the purple shaded regions in Fig. 2 indicate the advantage of our criteria over the method of Ref. \cite{Pryde}. In this respect, two approaches are complementary to each other in detecting steering of Gaussian states for which Gaussian criterion fails. 

\begin{figure}[!t]
  \centering
  \includegraphics[width=0.3\textwidth]{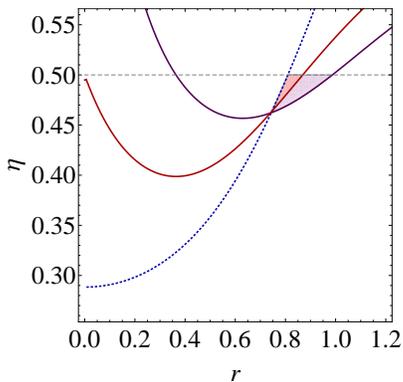}
\caption{Comparing the steering criterion in \cite{Pryde} with ours for the detection of two-mode squeezed states with squeezing $r$ under a loss channel with transmittance $\eta$. The mode $B$ undergoes the loss channel and the Gaussian states above each curve are detected for the steering from $B$ to $A$. Blue dotted: criterion of Ref. \cite{Pryde}, Red solid: 2-level TLOO, Purple solid: 3-level TLOO.}
\label{model}
\end{figure}

\section*{Methods}
To calculate the expectation values of the observables in our steering criteria, we need the density matrix elements in Fock basis, i.e., $ \rho_{m_1m_2n_1n_2}\equiv{\rm Tr} \{\rho | m_1 \rangle_A \langle n_1| \otimes | m_2 \rangle_B  \langle n_2 | \}$. In particular, a two mode squeezed state (squeezing: $r$), after the mode $B$ undergoes a vacuum noisy channel with transmittance $\eta$, gives  
\begin{equation}
\begin{array}{lllllllllll}
\rho_{0000} = \frac{2}{\cosh{(2r)}+1}, \\
 \rho_{0011}=\rho_{1100}=\frac{2 \sqrt{\eta} \sinh{(2r)}}{\left( \cosh{(2r)}+1 \right)^2}, \\
 \rho_{0022}=\rho_{2200} =\frac{2\eta\left( \cosh{(2r)} -1 \right)}{\left( \cosh{(2r)} +1 \right)^2}, \\
  \rho_{1010}=\frac{2 \left(1- \eta\right) \left( \cosh{(2r)}-1 \right)}{\left( \cosh{(2r)}+1 \right)^2}, \\
 \rho_{1021}=\rho_{2110}=\frac{2\sqrt{2}\left( 1- \eta\right) \sinh{(2r)}\left( \cosh{(2r)}-1 \right) }{\left( \cosh{(2r)}+1 \right)^3} ,  \\
\rho_{1111}=\frac{2 \eta \left( \cosh{(2r)} -1 \right)}{\left( \cosh{(2r)}+1 \right)^2}, \\
\rho_{1122}=\rho_{2211}= \frac{2\eta^{3/2} \sinh^3{(2r)}}{\left( \cosh{(2r)}+1 \right)^4}, \\
 \rho_{2020}=\frac{2\left( 1- \eta \right)^2 \left( \cosh{(2r)}-1 \right)^2}{\left( \cosh{(2r)}+1 \right)^3}, \\
 \rho_{2121} = \frac{4\eta(1-\eta) \left( \cosh{(2r)}-1 \right)^2}{\left( \cosh{(2r)}+1 \right)^3},\\
\rho_{2222}=\frac{2\eta^2 \left( \cosh{(2r)} -1 \right)^2}{\left( \cosh{(2r)}+1 \right)^3}, 
  \end{array}
\end{equation}
while other terms are zero in the basis of $ \left\{ | 0 \rangle, | 1 \rangle,  | 2 \rangle \right\}$.  

On the other hand, each single mode state in mode $A$ and $B$ is a thermal state, whose expectation values are also necessary to test our criterion. The single-mode thermal states are all diagonal in Fock basis, and the nonzero expectation values for the mode $B$ are given by 
\begin{equation}
\begin{array}{lll}
\langle B_0 \rangle_B = \langle | 0 \rangle \langle 0| \rangle_B = \frac{1}{1+\bar{n}} =\frac{2}{\eta \cosh{(2r)}-\eta+2}, \\
\langle B_1 \rangle_B = \langle | 1 \rangle \langle 1| \rangle_B = \frac{\bar{n}}{\left(1+\bar{n} \right)^2} = \frac{2\eta\left( \cosh{(2r)}-1 \right)}{\left( \eta\sinh{(2r)}-\eta+2 \right)^2}, \\
 \langle B_2 \rangle_B = \langle | 2 \rangle \langle 2| \rangle_B = \frac{\bar{n}^2}{\left(1+\bar{n} \right)^3}=\frac{2\eta^2\left( \cosh{(2r)}-1 \right)^2}{\left( \eta\sinh{(2r)}-\eta+2 \right)^3}, \\ \end{array}
\end{equation}
where $ \bar{n}$ is the mean photon number of the initial thermal state. For mode $A$ which does not undergo a lossy channel, we simply set $\eta=1$ in the above expressions.

\section*{Acknowledgements}
This work was supported by an NPRP grant 7-210-1-032 from Qatar National Research Fund.

\onecolumngrid
\appendix*
\setcounter{equation}{0}
\section{{\Large Supplemental Material}}
\subsection{A1. Proofs}
{\bf (i) Proof of Lemma}:
First, a direct calculation readily gives $ \sum_j \left(A_j^{(n)}\right)^2 = n \openone_n $. On the other hand, the sum of squares of expectation values is given by 
\begin{equation}
\begin{array}{llllll}
 \sum_j \langle A_j^{(n)} \rangle^2  & & =   \sum_{k=0}^{n-1} \langle |k \rangle \langle k | \rangle^2 \\
&  + & \frac{1}{2}\sum_{k<l}  \left( \langle |k \rangle\langle l | \rangle^2 + \langle |l \rangle\langle k | \rangle^2 + 2| \langle |k \rangle \langle l | \rangle |^2 \right) \\
&  - &\frac{1}{2} \sum_{k<l} \left( \langle |k \rangle\langle l | \rangle^2 + \langle |l \rangle\langle k | \rangle^2 - 2| \langle |k \rangle \langle l | \rangle |^2 \right) \\ 
&  &=  \sum_{k=0}^{n-1} \langle |k \rangle \langle k | \rangle^2 + 2 \sum_{k<l}| \langle |k \rangle \langle l | \rangle |^2 \\
&  &\leq  \sum_{k=0}^{n-1} \langle |k \rangle \langle k | \rangle^2 + 2 \sum_{k<l} \langle |k \rangle \langle k | \rangle\langle |l \rangle\langle l | \rangle \\
& & =    \langle \sum_{k=0}^{n-1}  |k \rangle\langle k | \rangle^2    \leq \langle \openone_n \rangle,
\end{array}
\end{equation} 
where we use the Cauchy inequality to obtain the inequality in the fifth line. The positivity of the variance and $ \openone_n^2 =\openone_n$ yield  the last inequality, which in turn gives the uncertainty relation in equation (8) of main text. The equality holds when a given state is a pure state within the Fock space spanned by $ \left\{ |0\rangle, ..., | n-1 \rangle \right\}$. 

{\bf (ii) Proof of Theorem}: As shown in \cite{Cavalcanti09}, if the set of observables satisfies uncertainty relations in a sum or product form as Eq. (8) of main text, the correlation of a bipartite quantum state described by the LHS models must satisfy a non-steerablity inequality in a form  
\begin{equation}
\label{infer}
\sum_{j} \delta_{inf}^{2} \left( A_j^{(n)} \right) \geq \left( n-1 \right) \langle \openone_n^{A} \rangle,
\end{equation}
where $ \delta_{inf}^{2}\left( A_j^{(n)} \right) $ is an inferred variance \cite{Reid89} defined by
\begin{equation}
\label{infervar}
\delta^2_{inf}\left(A_j^{(n)}\right)=\langle \left[ A_j^{(n)} -A_{j,est}^{(n)}\left( B_j^{(n')} \right) \right]^2 \rangle.
\end{equation}
$A_{j,est}^{(n)}\left(B_j^{(n')}\right)$ is an estimate based on Bob's outcome $ B_j^{(n')}$ \cite{Cavalcanti09, Reid89}. With a choice of linear estimate $ A_{j,est}^{(n)}\left( B_j^{(n')} \right)= -g_j B_j^{(n')}+ \langle A_j^{(n)} + g_j B_j^{(n')} \rangle$ ($g_j$ is an arbitrary real number) \cite{Cavalcanti09, Reid09} and setting $ g=g_1=g_2=...=g_n $, we obtain the non-steerability criterion in equation (9) of main text.

{\bf (iii) Proof of Proposition}: 
Similar to the analysis in \cite{Ji15}, let us first choose $g$ to make the left-hand side of equation (9) of main text as small as possible, i.e. 
\begin{equation}\
\label{set-g}
g=-\frac{\sum_j \langle A_j^{(n)} \otimes B_j^{(n')} \rangle -\langle A_j^{(n)} \rangle\langle B_j^{(n')} \rangle  }{\sum_j \delta^2\left(B_j^{(n')} \right)}.
\end{equation}
Plugging it to the inequality (9) of main text, we obtain the nonsteerability condition as
\begin{equation}
\label{steering1}
\begin{array}{ll}
| \sum_j \langle A_j^{(n)} \otimes B_j^{(n')} \rangle -\langle A_j^{(n)} \rangle \langle B_j^{(n')} \rangle |  \\
\leq \sqrt{ \left( \langle \openone_n^{A} \rangle -\sum_j \langle A_j^{(n)} \rangle^2 \right) \left( n' \langle \openone_{n'}^B \rangle - \sum_j \langle B_j^{(n')} \rangle^2 \right)}.
\end{array}
\end{equation}
Note that the left-hand side of Eq. (\ref{steering1}) has only the diagonal elements of correlation matrix. We may concentrate the correlation information onto the diagonal terms by taking a singular value decomposition of the correlation matrix $C_{nn'}^{TLOOs}$ ( $n^2 \times n'^2$ real matrix) using certain orthogonal matrices $ O_n^{A}$ and $O_{n'}^{B}$. Under this transformation, TLOOs remain TLOOs as already shown and the left-hand side of Eq. (\ref{steering1}) becomes a trace norm of the new correlation matrix. 
Using also the invariance of sum of squares of expectation values of TLOOs, the inequality (\ref{steering1}) corresponds to the inequality (10) of main text. 

\subsection{{A2. Expectation values of observables in Fock space} } 

In order to calculate the expectation values of Fock-basis observables for a two-mode Gaussian state, the multivariable Hermite polynomials are very useful \cite{Tatham14}. The multivariable (4 variables in our case) Hermite polynomials are given by 

\begin{equation}
\label{hermite}
H^{\left\{R,\theta\right\}}_{m_1,m_2,n_1,n_2} \left( y_1, y_2, y_3, y_4 \right)=(-1)^{m_1+m_2+n_1+n_2}\exp\left[ \vec{y}^T R \vec{y} \right] \frac{\partial^{m_1}}{\partial y^{m_1}_1}\frac{\partial^{m_2}}{\partial y^{m_2}_2}\frac{\partial^{n_1}}{\partial y^{n_1}_3}\frac{\partial^{n_2}}{\partial y^{n_2}_4} \exp\left[ -\vec{y}^T R \vec{y}-\theta\vec{y}\right],
\end{equation}

where $ \vec{y}= \left( y_1, y_2, y_3, y_4 \right)^T$. The matrix $R$ is a $ 4 \times 4$ matrix related to the covariance matrix of the two-mode Gaussian state. With these Hermite polynomials, the matrix elements in Fock basis are given by \cite{Tatham14}
\begin{equation}
\label{fockbasis}
_A\langle m_1 | _B\langle m_2 | \rho_{AB} | n_1 \rangle_A | n_2 \rangle_B =\frac{4 H^{\left\{R,0 \right\}}_{m_1,m_2,n_1,n_2}\left(0,0,0,0\right)}{\sqrt{m_1 ! m_2 ! n_1 ! n_2 !} \sqrt{ \det{\left(\gamma_{AB}+ \openone \right)}}},
\end{equation} 
where $R=BU\left[ \left(\gamma_{AB} + \openone \right)^{-1} -\frac{1}{2}\openone \right] U^{\dagger} D$ and 

\begin{equation}
U=\frac{1}{\sqrt{2}} \left( \begin{array}{cccc}
1 & i & 0 & 0 \\
1 & -i & 0 & 0 \\
0 & 0 & 1 & i \\
0 & 0 & 1 & -i\\
\end{array} \right) ,\;\; B=\left( \begin{array}{cccc} 
1 & 0 & 0 & 0 \\
0 & 0 & 1 & 0 \\
0 & 1 & 0 & 0 \\
0 & 0 & 0 &1 \\ 
\end{array} \right) ,\;\; D= \left( \begin{array}{cccc} 
0 & 0 & 1 & 0 \\
1 & 0 & 0 & 0 \\
0 & 0 & 0 & 1 \\
0 & 1 & 0 & 0 \\
\end{array} \right) .
\end{equation}

Now let us consider a covariance matrix $\gamma_{AB}$ in a standard form
\begin{equation}
\label{bscova}
\gamma_{AB} = \left( \begin{array}{cccc}
 a & 0 & c_1 & 0 \\ 
 0 & a & 0 & - c_2  \\ 
 c_1 & 0 & b & 0 \\ 
 0 & - c_2 & 0 & b \\ 
 \end{array} \right) .
\end{equation} 
In this case the $R$ matrix in equation (\ref{fockbasis}) is given by 
\begin{equation}
\label{matrixr}
R=\frac{1}{2}\left( \begin{array}{cccc}
\tilde{a}_1 & \tilde{c}_1 & \tilde{a}_2 & \tilde{c}_2 \\
\tilde{c}_1 & \tilde{b}_1 & \tilde{c}_2 & \tilde{b}_2 \\
\tilde{a}_2 & \tilde{c}_2 & \tilde{a}_1 & \tilde{c}_1 \\
\tilde{c}_2 & \tilde{b}_2 & \tilde{c}_1 & \tilde{b}_1 \\
\end{array} \right),
\end{equation}
where 
\begin{equation}
\begin{array}{cccccc}
\tilde{a}_1=\frac{\left(b+1\right)\left(c_1^2-c_2^2 \right)}{\left[ \left(a+1\right)\left(b+1\right)-c_1^2\right]\left[ \left(a+1\right)\left(b+1\right)-c_2^2 \right]} , \\
\tilde{a}_2=-1+\frac{\left(b+1\right)\left[2\left(a+1\right)\left(b+1\right)-\left(c_1^2+c_2^2\right)\right]}{\left[ \left(a+1\right)\left(b+1\right)-c_1^2\right]\left[ \left(a+1\right)\left(b+1\right)-c_2^2 \right]}, \\
\tilde{b}_1=\frac{\left(a+1\right)\left(c_1^2-c_2^2\right)}{\left[ \left(a+1\right)\left(b+1\right)-c_1^2\right]\left[ \left(a+1\right)\left(b+1\right)-c_2^2 \right]} , \\
\tilde{b}_2=-1+\frac{\left(a+1\right)\left[2\left(a+1\right)\left(b+1\right)-\left(c_1^2+c_2^2\right)\right]}{\left[ \left(a+1\right)\left(b+1\right)-c_1^2\right]\left[ \left(a+1\right)\left(b+1\right)-c_2^2 \right]}, \\
\tilde{c}_1=\frac{-\left[\left(a+1\right)\left(b+1\right)-c_1c_2\right]\left(c_1+c_2\right)}{\left[ \left(a+1\right)\left(b+1\right)-c_1^2\right]\left[ \left(a+1\right)\left(b+1\right)-c_2^2 \right]} , \\
\tilde{c}_2=\frac{-\left[\left(a+1\right)\left(b+1\right)+c_1c_2\right]\left(c_1-c_2\right)}{\left[ \left(a+1\right)\left(b+1\right)-c_1^2\right]\left[ \left(a+1\right)\left(b+1\right)-c_2^2 \right]}. \\
\end{array}
\end{equation}
For the covariance matrix $\gamma_{AB}^{TMSV}$ of a two-mode squeezed vacuum state (TMSV) with squeezing parameter $r$, the matrix $R^{TMSV}$ is given in a simple form

\begin{equation}
\label{matrxirtmsv}
R^{TMSV}=\frac{1}{2}\left( \begin{array}{cccc}
0 & -\frac{\sinh{2r}}{\cosh{2r}+1} & 0 & 0 \\
-\frac{\sinh{2r}}{\cosh{2r}+1} & 0 & 0 & 0 \\
0 & 0 & 0 & -\frac{\sinh{2r}}{\cosh{2r}+1} \\
0 & 0 & -\frac{\sinh{2r}}{\cosh{2r}+1} & 0 \\
\end{array} \right)
\end{equation}

If this TMSV goes through loss in mode $B$ only, the matrix $R^{TMSV}_{LB}$ is given by

\begin{equation}
\label{matrixrloss}
R^{TMSV}_{LB}=\frac{1}{2}\left( \begin{array}{cccc}
0 & -\frac{\sqrt{\eta}\sinh{2r}}{\cosh{2r}+1} & -\frac{\left(1-\eta\right)\left( \cosh{2r}-1\right)}{\cosh{2r}+1} & 0 \\
-\frac{\sqrt{\eta}\sinh{2r}}{\cosh{2r}+1} & 0 & 0 & 0 \\
-\frac{\left(1-\eta\right)\left( \cosh{2r}-1\right)}{\cosh{2r}+1} & 0 & 0 & -\frac{\sqrt{\eta}\sinh{2r}}{\cosh{2r}+1} \\
0 & 0 & -\frac{\sqrt{\eta}\sinh{2r}}{\cosh{2r}+1} & 0 \\
\end{array} \right),
\end{equation}

with $\eta$ transmittance rate. We can similarly obtain the matrix $R^{TMSV}_{LG}$ for the case of amplification. \\
 
Therefore, using the matrix elements in equation (\ref{fockbasis}), we can calculate the expectation values of any observables $ A_k^{(n)} \otimes B_l^{(n')}$ to test our criterion in the main text. 

\subsection{{A3. Singular value decomposition of correlation matrix}}
We here show that violation of the steering criterion inequality (10) of the main text is equivalent to the violation of equation (9) in the main text. In view of {\bf Proposition}, let us assume that a given two-mode state $\rho_{AB}$ violates the inequality 

\begin{equation}
\label{Asteering-tr}
\| C^{TLOOs}_{nn'}\|_{tr} \leq \sqrt{\left(\langle \openone_n^A \rangle -\sum_j \langle A_j^{(n)} \rangle^2 \right) \left( n'\langle \openone_{n'}^B \rangle -\sum_j \langle B_j^{(n')} \rangle^2\right)}. 
\end{equation} 
  
This violation means that there exists a set of TLOOs $\left\{ \tilde{A}_j^{(n)} \right\}$, $\left\{ \tilde{B}_j^{(n')} \right\}$, which satisfies an inequality 

\begin{equation}
\label{singular1}
\sum_j \langle \tilde{A}_j^{(n)} \otimes \tilde{B}_j^{(n')} \rangle -\langle \tilde{A}_j^{(n)} \rangle \langle \tilde{B}_j^{(n')} \rangle > \sqrt{\left(\langle \openone_n^A \rangle -\sum_j \langle A_j^{(n)} \rangle^2 \right) \left( n'\langle \openone_{n'}^B \rangle -\sum_j \langle B_j^{(n')} \rangle^2\right)}.
\end{equation}
  
The sets $\left\{ \tilde{A}_j^{(n)} \right\}$ and $\left\{ \tilde{B}_j^{(n')} \right\}$ can be chosen by employing the eigenvectors of $ C^{TLOOs}_{nn'} \left(C^{TLOOs}_{nn'}\right)^T$ and $\left(C^{LOOs}_{nn'}\right)^T C^{LOOs}_{nn'}$ used for the singular value decomposition
\begin{equation}
\label{TOS}
O_n^A C^{TLOOs}_{nn'} \left(O_{n'}^B\right)^{T} = C^{TLOOs}_{nn',\,singular}.
\end{equation}
Here $C^{TLOOs}_{nn',\,singular}$ is a diagonal matrix with non-negative elements, and $O^A_n$ and $O^B_{n'}$ are truncated orthogonal matrices of which columns are eigenvectors of $ C^{TLOOs}_{nn'} \left(C^{TLOOs}_{nn'}\right)^T$ and $\left(C^{TLOOs}_{nn'}\right)^T C^{TLOOs}_{nn'}$, respectively \cite{Horn2}.  \\
\\
Now, let us consider steering criterion in equation (9) of {\bf Theorem} and equation (A.4) with the observables in equation (\ref{singular1}). We can then derive the following 
 
\begin{equation}
\begin{array}{llll}
\sum_k^N \delta^{2} \left(  \tilde{A}_k^{(n)} \otimes \openone + g \openone \otimes \tilde{B}_k^{(n')} \right) 
& =  \sum_{k}  \delta^2 \left( \tilde{A}_k^{(n)} \right) + g^2 \sum_k \delta^2 \left( \tilde{B}_k^{(n')} \right) + 2g \sum_k  \left( \langle \tilde{A}_k \otimes \tilde{B}_k \rangle -\langle \tilde{A}_k \rangle \langle \tilde{B}_k \rangle \right)  \\
& = \frac{-\left( \sum_k \langle \tilde{A}_k^{(n)} \otimes \tilde{B}_k^{(n')} \rangle -\langle \tilde{A}_k^{(n)} \rangle \langle  \tilde{B}_k^{(n')} \rangle \right)^2 }{\sum_k \delta^2\left(\tilde{B}_k^{(n')}\right)}+\sum_k \delta^2\left( \tilde{A}_k^{(n)} \right) \\
& <\frac{\left(n'\langle \openone^{B}_{(n')} \rangle -\sum_k\langle \tilde{B}_k^{(n')} \rangle^2\right) \left(  \sum_k \langle A_k^{(n)} \rangle^2  -\langle \openone_n^{A} \rangle \right)}{n'\langle \openone^{B}_{(n')} \rangle -\sum_k\langle \tilde{B}_k^{(n)} \rangle^2} + n\langle \openone_n^{A} \rangle-  \sum_k \langle A_k^{(n)} \rangle^2  \\
& = \left(n-1\right) \langle \openone_n^A \rangle, 
\end{array}
\end{equation}

where we used optimal $ g=-\frac{\sum_{k}\langle \tilde{A}_k^{(n)} \otimes \tilde{B}_k^{(n')} \rangle -\langle \tilde{A}_k^{(n)} \rangle\langle \tilde{B}_k^{(n')} \rangle}{\sum_{k} \delta^2 \left(\tilde{B}_k^{(n')}\right)}$ as in equation (A.4) and $ \sum_{k} \delta^2 \left( \tilde{A}_k^{(n)} \right)=n\langle \openone_n^{A} \rangle -\sum_k \langle \tilde{A}_k^{(n)} \rangle^2$, $\sum_{k} \delta^2 \left( \tilde{B}_k^{(n')} \right)= n'\langle \openone_{n'}^{B} \rangle -\sum_k \langle \tilde{B}_k^{(n')} \rangle^2$. The inequality in the third line is given by equation (\ref{singular1}) which we assumed. In summary, the violation of equation (\ref{Asteering-tr}) with TLOOs which are constructed by singular value decomposition in equation (\ref{TOS}) is equivalent to the violation of the steering criterion in equation (9) of {\bf Theorem} with the same TLOOs.

\end{document}